\documentclass[aps,prb,showpacs,twocolumn]{revtex4-1}
\usepackage{amsfonts}
\usepackage{amssymb}
\usepackage{amsmath}
\usepackage{graphicx}
\usepackage{epsfig}

\begin{document}

\title{Non-Hermitian anisotropic $XY$ model with intrinsic rotation-time
reversal symmetry}
\author{X. Z. Zhang and Z. Song}
\email{songtc@nankai.edu.cn}
\affiliation{School of Physics, Nankai University, Tianjin 300071, China}

\begin{abstract}
We systematically study the non-Hermitian version of the one-dimensional
anisotropic $XY$ model, which in its original form, is a unique exactly
solvable quantum spin model for understanding the quantum phase transition.
The distinguishing features of this model are that it has full real spectrum
if all the eigenvectors are intrinsic
rotation-time reversal ($\mathcal{RT}$)symmetric rather than parity-time
reversal ($\mathcal{PT}$)symmetric, and that its Hermitian counterpart
is shown approximately to be an experimentally accessible system, an
isotropic $XY$ spin chain with nearest neighbor coupling. Based on the exact
solution, exceptional points which separated the unbroken and broken
symmetry regions are obtained and lie on a hyperbola in the thermodynamic
limit. It provides a nice paradigm to elucidate the complex quantum
mechanics theory for a quantum spin system.
\end{abstract}

\pacs{75.10.Jm, 03.65.-w, 11.30.Er, 64.70.Tg}
\maketitle


\section{Introduction}

\label{sec_intro} In recent years, much effort has been devoted to establish
a parity-time ($\mathcal{PT}$) symmetric quantum theory as a complex
extension of the conventional quantum mechanics \cite{Bender 98,Bender
99,Dorey 01,Bender 02,A.M43,A.M37,A.M36,Jones} since the seminal discovery
by Bender \cite{Bender 98}. A cornerstone of the theory is the fact that the
non-Hermitian Hamiltonian with $\mathcal{PT}$ symmetry can have an entirely
real energy spectrum. It is also motivated by the interest in complex
potentials in both theoretical and experimental aspects, since the imaginary
potential could be realized by complex index in optics \cite%
{Bendix,Joglekar,Keya,YDChong,LonghiLaser}. A natural question to ask is
whether the non-Hermitian quantum theory prefers to the $\mathcal{PT}$
symmetry. While there is as yet no answer to this question, we can gain some
insight regarding the pseudo-Hermiticity of a non-Hermitian model without
the $\mathcal{PT}$ symmetry. On the other hand, the non-Hermitian quantum
model in discrete system is a nice testing ground for the study of the
non-Hermitian quantum mechanics because of its analytical and numerical
tractability. As simplified models, tight-binding and quantum spin models
can capture the essential features of many discrete systems. In general, the
non-Hermiticity arises from $\mathcal{PT}$ symmetric on-site imaginary
potentials in tight-binding models \cite%
{MZnojil,Bendix,Longhi,Joglekar,Joglekar83,Weston,Fring,Ghosh,Jin
Liang,ZLi,Bousmina,Ozlem,Fabio,HWH}, and imaginary magnetic fields in
quantum spin models \cite{Giorgi,X. Z. Zhang}.\ In addition, the complex
coupling constant\ between spins can also introduce the non-Hermiticity,
which relates to the spin, the intrinsic degree of freedom. Then the spin
rotation may take the role of the parity operation for constructing a
pseudo-Hermitian spin system.

In this paper, we propose a pseudo-Hermitian model without $\mathcal{PT}$
symmetry\ explicitly, but with intrinsic rotation-time ($\mathcal{RT}$)
reversal symmetry. It is a non-Hermitian version of the one-dimensional
anisotropic $XY$ model. The original Hermitian $XY$ model was initially
solved and has became the paradigm of the quantum spin system possessing a
second-order quantum phase transition \cite{E. Lieb}. We will show that this
model has full real spectrum if all the eigenvectors have $\mathcal{RT}$
rather than parity-time reversal ($\mathcal{PT}$) symmetry, and its
Hermitian counterpart is an isotropic $XY$\ spin chain. The result for such
a concrete example may have profound theoretical and methodological
implications.

This paper is organized as follows. In Section \ref{sec_model}, we present
the model Hamiltonian. Based on the solutions we investigate the phase
diagram and analyze the symmetry of the ground state. In Section \ref{sec
Hermitian counterpart}, we construct the Hermitian counterpart of the model
and its approximate reduced form. Finally, we give a summary and discussion
in Section \ref{sec_summary}.

\section{Hamiltonian and intrinsic $\mathcal{RT}$\ symmetry}

\label{sec_model}The model Hamiltonian of the non-Hermitian anisotropic
one-dimensional spin-$\frac{1}{2}$ $XY$ model in a transverse magnetic field
$\lambda $ for $N$ particles is given by

\begin{equation}
H=J\sum\limits_{j=1}^{N}\left( \frac{1+i\gamma }{2}\sigma _{j}^{x}\sigma
_{j+1}^{x}+\frac{1-i\gamma }{2}\sigma _{j}^{y}\sigma _{j+1}^{y}+\lambda
\sigma _{j}^{z}\right)   \label{H}
\end{equation}%
$\sigma _{j}^{\alpha }$ ($\alpha =x,$ $y,$ $z$) are the Pauli operators on
site $j$, and satisfy the periodic boundary condition $\sigma _{j}^{\alpha
}\equiv \sigma _{j+N}^{\alpha }$. For the sake of simplicity, we only
concern the case of even $N$, the conclusion is available in the case of odd
$N$. In comparison with the model proposed by Giorgi \cite{Giorgi}, the
present model is not $\mathcal{PT}$ symmetric. The non-Hermiticity arises
from the imaginary anisotropic parameters\ $\pm i\gamma $. The Hermitian
version of the $XY$ model is completely solved by applying the Jordan-Wigner
\cite{P. Jordan}, Fourier and Bogoliubov transformations \cite{E. Lieb,P.
Pfeuty}. The Jordan-Wigner transformation maps the Pauli operators into
canonical fermions, while the Fourier transformation essentially decomposes
the Hamiltonian into invariant subspaces due to the translational symmetry
of the system. We will see that these transformations are applicable in
solving the Hamiltonian (\ref{H}) if we extend it to its complex versions.
Before solving the Hamiltonian, it is profitable to investigate the symmetry
of the system and its breaking in the eigenstates. By direct derivation, we
have $\left[ \mathcal{R},H\right] \neq 0$ and $\left[ \mathcal{T},H\right]
\neq 0$, but

\begin{equation}
\left[ \mathcal{RT},H\right] =0.  \label{RT symmetry}
\end{equation}%
The Hamiltonian is rotation-time ($\mathcal{RT}$) reversal invariant, where
the linear rotation operator $\mathcal{R}$ has the function of rotating each
spin by $\pi /2$ about the $z$-axis
\begin{equation}
\mathcal{R}\equiv \exp \left[ -i\left( \pi /4\right)
\sum\nolimits_{j=1}^{N}\sigma _{j}^{z}\right] ,  \label{R}
\end{equation}%
and the antilinear time reversal operator $\mathcal{T}$ has the function $%
\mathcal{T}i\mathcal{T=-}i$. Before we consider the general non-Hermitian $XY
$ model, we highlight key ideas on two limiting cases. When $\gamma =0$, the
Hamiltonian (\ref{H}) reduces to the ordinary $XY$ model with external field
in $z$ direction%
\begin{equation}
H_{0}=\frac{J}{2}\sum\limits_{j=1}^{N}\left( \sigma _{j}^{x}\sigma
_{j+1}^{x}+\sigma _{j}^{y}\sigma _{j+1}^{y}+2\lambda \sigma _{j}^{z}\right) ,
\end{equation}%
which has full real spectrum and its eigenstates can always be written as
the eigenstates of operator $\mathcal{RT}$. On the other hand, when $\gamma
\gg \lambda $ and $1$, the Hamiltonian (\ref{H}) reduces to
\begin{equation}
H_{\infty }=\frac{iJ\gamma }{2}\sum\limits_{j=1}^{N}\left( \sigma
_{j}^{x}\sigma _{j+1}^{x}-\sigma _{j}^{y}\sigma _{j+1}^{y}\right) ,
\end{equation}%
which has full imaginary spectrum. Then for an eigenstate $\left\vert \psi
_{n}\right\rangle $ of $H_{\infty }$\ with nonzero eigenvalue $E_{n}$, i.e.,
$H_{\infty }\left\vert \psi _{n}\right\rangle =E_{n}\left\vert \psi
_{n}\right\rangle ,$ we have%
\begin{equation}
H_{\infty }\mathcal{RT}\left\vert \psi _{n}\right\rangle =-E_{n}\mathcal{RT}%
\left\vert \psi _{n}\right\rangle
\end{equation}%
due to the facts of Eq. (\ref{RT symmetry}) and $E_{n}^{\ast }=-E_{n}$. The
eigenstate $\left\vert \psi _{n}\right\rangle $ obviously breaks the $%
\mathcal{RT}$\ symmetry. These results strongly imply that the $\mathcal{RT}$%
\ symmetry in the present model plays the same role as $\mathcal{PT}$\
symmetry\ in the $\mathcal{PT}$\ pseudo-Hermitian system. It motivates
further study of such a model systematically.
\begin{figure*}[tbp]
\includegraphics[ bb=41 176 541 618, width=5.5 cm, clip]{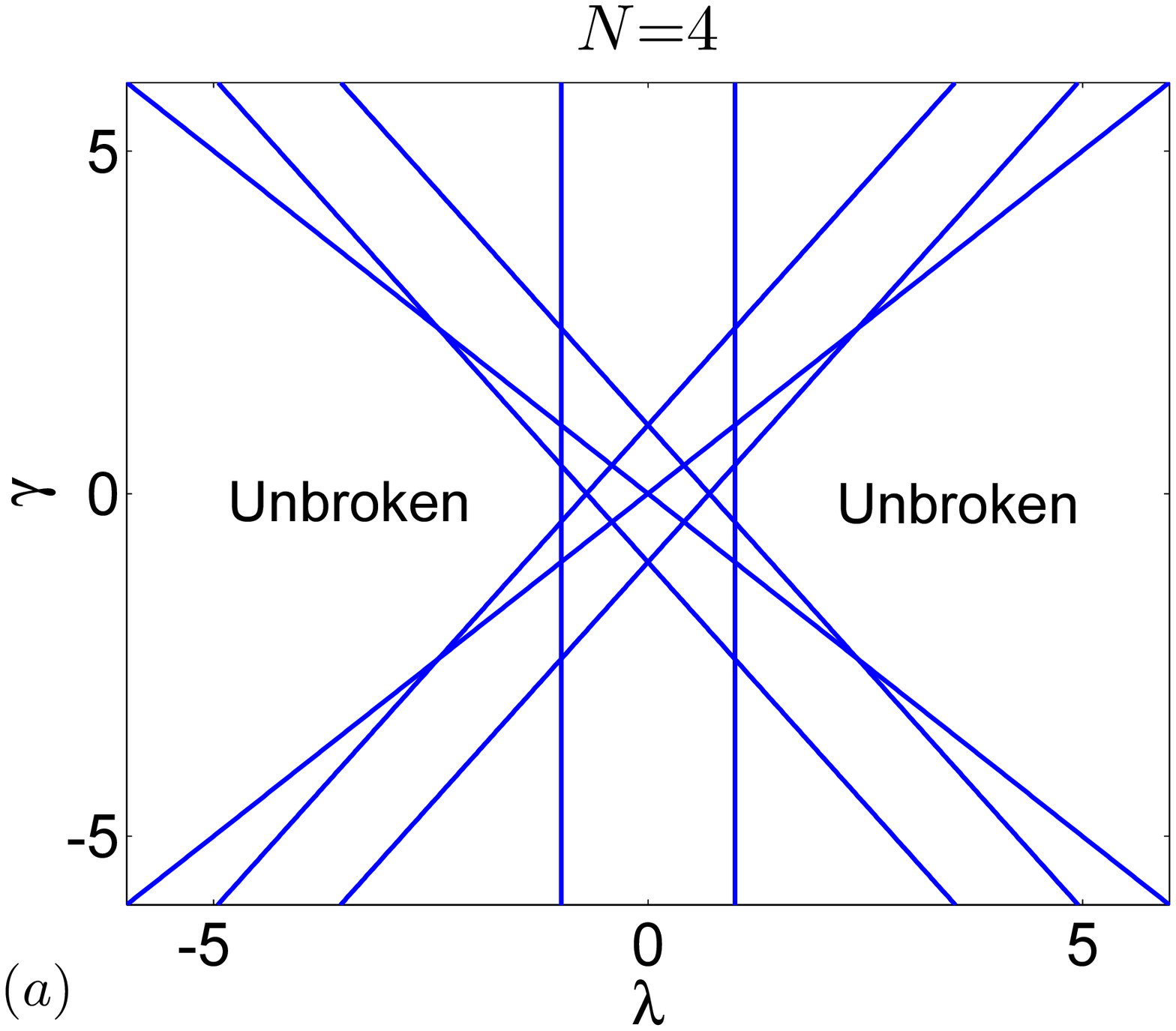} %
\includegraphics[ bb=41 176 541 618, width=5.5 cm, clip]{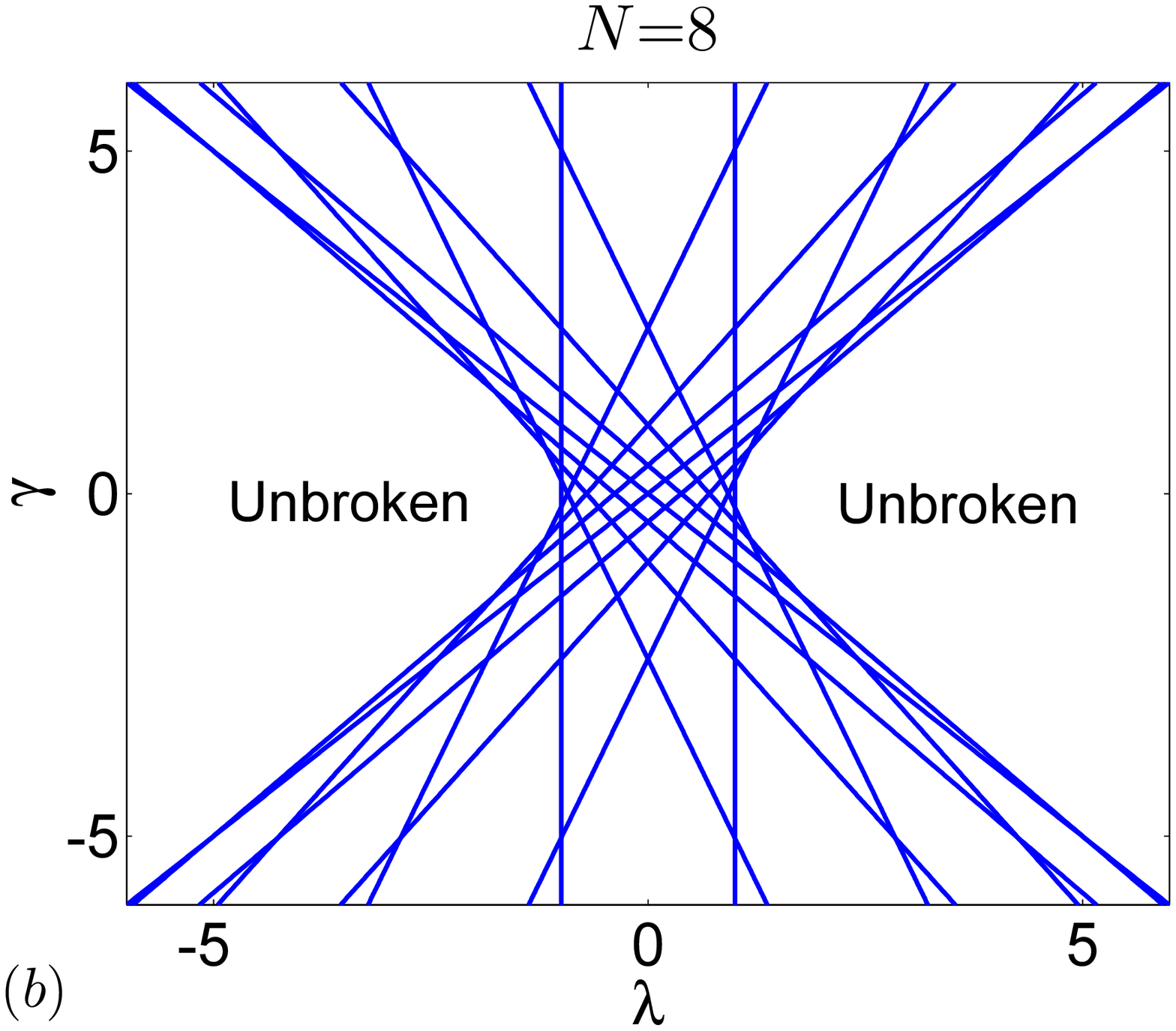} %
\includegraphics[ bb=41 176 541 618, width=5.5 cm, clip]{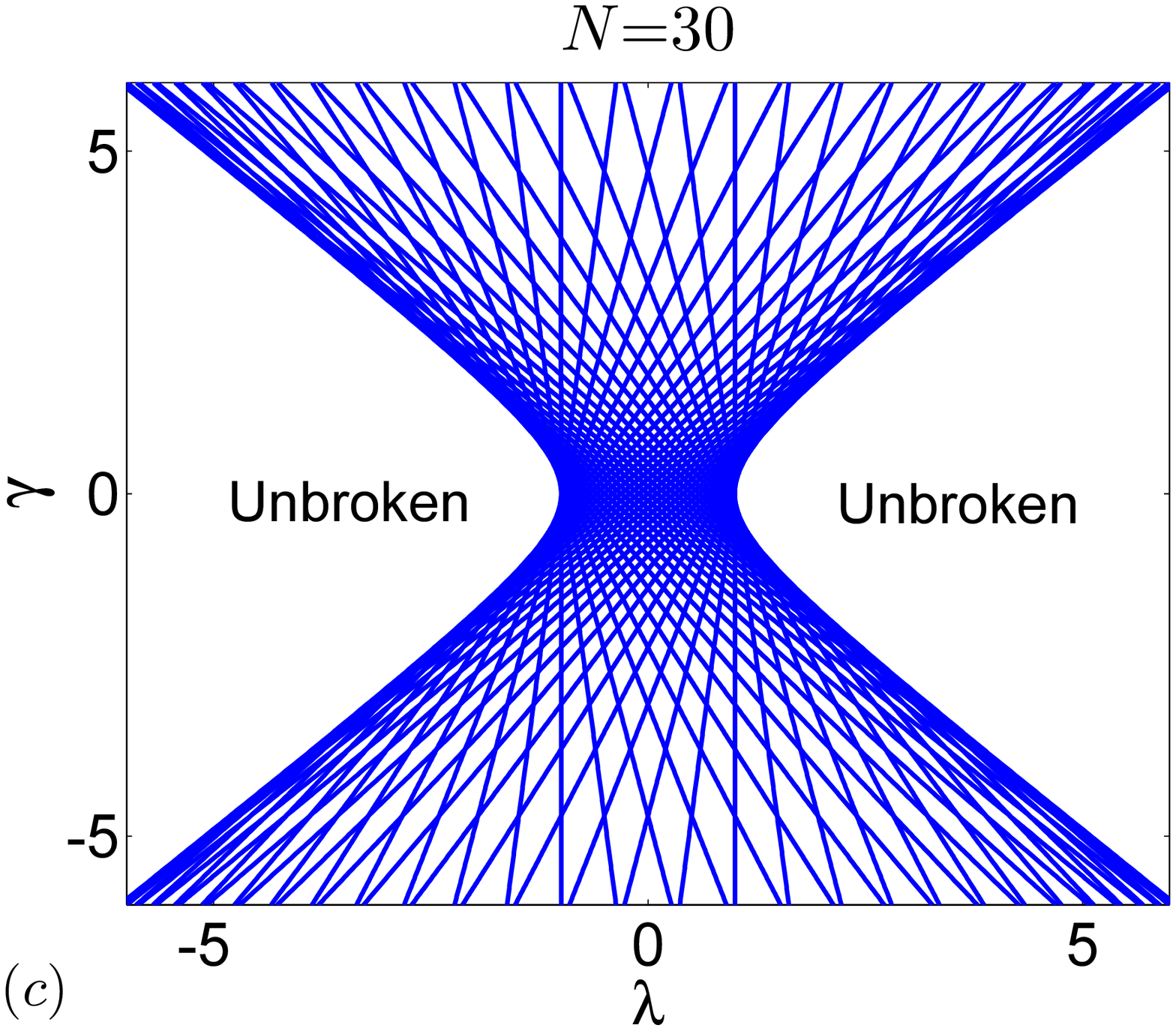}
\caption{(Color online) Plots of the phase diagrams for the systems with $N=4
$, $8$, and $30$, respectively. It is shown that as $N$ increases, the
boundary approaches to the hyperbola of Eq. (\protect\ref{hyperbola}). Note
that the broken region does not include the line $\protect\gamma =0$.}
\label{fig1}
\end{figure*}


\subsection{Solutions}

Now we consider the solution of the non-Hermitian $XY$ Hamiltonian of Eq. (%
\ref{H}). We note that the\textbf{\ }general solutions of the Hermitian
anisotropic $XY$ Hamiltonian are not restricted to the real anisotropic
parameter. We will show that in the case of imaginary parameter, eigenstates
and energies are\ still accessible. As the same procedures performed in
solving the Hermitian Hamiltonian, we take the Jordan-Wigner transformation
\cite{P. Jordan}%
\begin{eqnarray}
\sigma _{j}^{x} &=&-\prod\limits_{l<j}\left( 1-2c_{l}^{\dag }c_{l}\right)
\left( c_{j}^{\dag }+c_{j}\right) ,  \notag \\
\sigma _{j}^{y} &=&-i\prod\limits_{l<j}\left( 1-2c_{l}^{\dag }c_{l}\right)
\left( c_{j}^{\dag }-c_{j}\right) , \\
\sigma _{j}^{z} &=&1-2c_{j}^{\dag }c_{j},  \notag
\end{eqnarray}%
to replace the Pauli operators by the fermionic operators $c_{j}$. We note
that the parity of the number of fermions

\begin{equation}
\Pi =\prod_{l=1}^{N}\left( \sigma _{l}^{z}\right) =\left( -1\right) ^{N_{p}}
\end{equation}%
\bigskip is a conservative quantity, i.e., $\left[ H,\Pi \right] =0$, where $%
N_{p}=\sum_{j=1}^{N}c_{j}^{\dag }c_{j}$. Then the Hamiltonian (\ref{H}) can
be rewritten as

\begin{equation}
H=\sum_{\eta =+,-}P_{\eta }H_{\eta }P_{\eta },
\end{equation}%
where
\begin{equation}
P_{\eta }=\frac{1}{2}\left( 1+\eta \Pi \right)
\end{equation}%
is the projector on the subspaces with even ($\eta =+$) and odd ($\eta =-$) $%
N_{p}$. The Hamiltonian in each invariant subspaces has the form%
\begin{eqnarray}
H_{\eta } &=&J\sum\limits_{j=1}^{N-1}\left( c_{j}^{\dag
}c_{j+1}+c_{j+1}^{\dag }c_{j}+i\gamma c_{j}^{\dag }c_{j+1}^{\dag }+i\gamma
c_{j+1}c_{j}\right)  \notag \\
&&-\eta \left( c_{N}^{\dag }c_{1}+c_{1}^{\dag }c_{N}+i\gamma c_{N}^{\dag
}c_{1}^{\dag }+i\gamma c_{1}c_{N}\right) \\
&&-2J\lambda \sum\limits_{j=1}^{N}c_{j}^{\dag }c_{j}+NJ\lambda .  \notag
\end{eqnarray}%
Taking the Fourier transformation

\begin{equation}
c_{j}=\frac{1}{\sqrt{N}}\sum\limits_{k_{\pm }}e^{ik_{\pm }j}c_{k_{\pm }}
\end{equation}%
for the Hamiltonians $H_{\pm }$, we have

\begin{eqnarray}
H_{\eta } &=&-J\sum\limits_{k_{\eta }}[2\left( \lambda -\cos k_{\eta
}\right) c_{k_{\eta }}^{\dag }c_{k_{\eta }}  \label{H_eta} \\
&&-\gamma \sin k_{\eta }\left( c_{-k_{\eta }}c_{k_{\eta }}+c_{-k_{\eta
}}^{\dag }c_{k_{\eta }}^{\dag }\right) -\lambda ]  \notag
\end{eqnarray}%
where $k_{+}=2\pi \left( m+1/2\right) /N$, $k_{-}=2\pi m/N$, $%
m=0,1,2,...,N-1 $.

So far the procedures are the same as those for solving the Hermitian
version of $H$. To diagonalize the non-Hermitian Hamiltonian, we introduce
the Bogoliubov transformation in complex version:%
\begin{eqnarray}
A_{k_{\eta }} &=&\cos \left( \frac{\theta }{2}\right) c_{k_{\eta }}-i\sin
\left( \frac{\theta }{2}\right) c_{-k_{\eta }}^{\dag }  \label{AA} \\
\overline{A}_{k_{\eta }} &=&\cos \left( \frac{\theta }{2}\right) c_{k_{\eta
}}^{\dagger }+i\sin \left( \frac{\theta }{2}\right) c_{-k_{\eta }}  \notag
\end{eqnarray}%
where%
\begin{equation}
\tan \left( \theta \right) =\frac{i\gamma \sin k_{\eta }}{\left( \lambda
-\cos k_{\eta }\right) }.
\end{equation}%
We would like to point out that this is the crucial step to solve the
non-Hermitian Hamiltonian, which essentially establish the biorthogonal
bases. Obviously, complex Bogoliubov modes $\left( A_{k_{\eta }},\overline{A}%
_{k_{\eta }}\right) $ satisfy the canonical commutation relations%
\begin{eqnarray}
\left\{ A_{k_{\eta }},\overline{A}_{k_{\eta }^{\prime }}\right\}  &=&\delta
_{k_{\eta },k_{\eta }^{\prime }},  \label{canon} \\
\left\{ A_{k_{\eta }},A_{k_{\eta }^{\prime }}\right\}  &=&\left\{ \overline{A%
}_{k_{\eta }},\overline{A}_{k_{\eta }^{\prime }}\right\} =0;  \notag
\end{eqnarray}%
which result in the diagonal form of the Hamiltonian%
\begin{equation}
H_{\eta }=\sum\limits_{k_{\eta }}\epsilon \left( \lambda ,k_{\eta },\gamma
\right) \left( \overline{A}_{k_{\eta }}A_{k_{\eta }}-\frac{1}{2}\right) .
\label{H_+/-}
\end{equation}%
Here the single-particle spectrum in each subspace is%
\begin{equation}
\epsilon \left( \lambda ,k_{\eta },\gamma \right) =-2J\sqrt{\left( \lambda
-\cos k_{\eta }\right) ^{2}-\gamma ^{2}\sin ^{2}k_{\eta }}.
\end{equation}%
Note that the Hamiltonian $H_{\eta }$\ is still non-Hermitian due to the
fact that $\overline{A}_{k_{\eta }}\neq A_{k_{\eta }}^{\dag }$. Accordingly,
the eigenstates of $H_{\eta }$ can be written as the form%
\begin{equation}
\prod\limits_{\left\{ k_{\eta }\right\} }\overline{A}_{k_{\eta }}\left\vert
\text{G}_{\eta }\right\rangle ,
\end{equation}%
which constructs the biorthogonal set associated with the eigenstates%
\begin{equation}
\left\langle \text{G}_{\eta }\right\vert \prod\limits_{\left\{ k_{\eta
}\right\} }A_{k_{\eta }}
\end{equation}%
of the Hamiltonian $H_{\eta }^{\dag }$, where
\begin{equation}
\left\vert \text{G}_{\eta }\right\rangle =\prod_{k>0}\left[ \cos \left(
\frac{\theta }{2}\right) +i\sin \left( \frac{\theta }{2}\right)
c_{k}^{\dagger }c_{-k}^{\dagger }\right] \left\vert \text{Vac}\right\rangle
\label{GS}
\end{equation}%
is the ground state of $H_{\eta }$, and $\left\vert \text{Vac}\right\rangle $%
\ is the vacuum state of the fermion $c_{j}$. In the following, we will
investigate the phase diagram based on the properties of the solutions.

\subsection{Phase diagram}

It is clear that when any one of the momentum $k_{\eta }$\ satisfies%
\begin{equation}
\left\vert \lambda -\cos k_{\eta }\right\vert <\left\vert \gamma \sin
k_{\eta }\right\vert ,
\end{equation}%
the imaginary energy level appears in single-particle spectrum, which leads
to the occurrence of complex energy level for the Hamiltonian (\ref{H}), and
the $\mathcal{RT}$ symmetry is broken in the corresponding eigenstates. This
can be seen from the properties of the single-particle spectrum and the
ground states $\left\vert \text{G}_{\eta }\right\rangle $.

Firstly, we focus on the boundary between the broken and unbroken symmetry
regions. There are totally $2N$ equations in the form of $\left\vert \lambda
-\cos k_{\eta }\right\vert =\left\vert \gamma \sin k_{\eta }\right\vert $
for all the possible value of $k_{\eta }$, only one of\ which determines the
line along the boundary of the diagram within certain region in the
parameters $\lambda $ and $\gamma $ plane. Then the phase boundary is
dog-leg path for finite $N$, and becomes a smooth loop for the infinite $N$.
In the thermodynamic limit, the momentum $k_{\eta }$\ becomes continuous,
then the phase boundary as a curve can be given by the following set of
parametric equations \cite{Jin Liang}%
\begin{eqnarray}
\frac{\partial \epsilon \left( \lambda _{c},k_{\eta },\gamma _{c}\right) }{%
\partial k_{\eta }} &=&0, \\
\epsilon \left( \lambda _{c},k_{\eta },\gamma _{c}\right)  &=&0.
\end{eqnarray}%
Straightforward algebra gives the analytical boundary curve as%
\begin{equation}
\lambda _{c}^{2}-\gamma _{c}^{2}=1,  \label{hyperbola}
\end{equation}%
which is a hyperbola. In addition, the broken region does not include the
line $\gamma =0$. In Fig. \ref{fig1}, we plot the phase diagrams for the
systems with $N=4$, $8$, and $30$, respectively. It is shown that as $N$
increases, the boundary approaches to the hyperbola of Eq. (\ref{hyperbola}%
). 
\begin{figure*}[tbp]
\includegraphics[ bb=24 178 542 584, width=6.5 cm, clip]{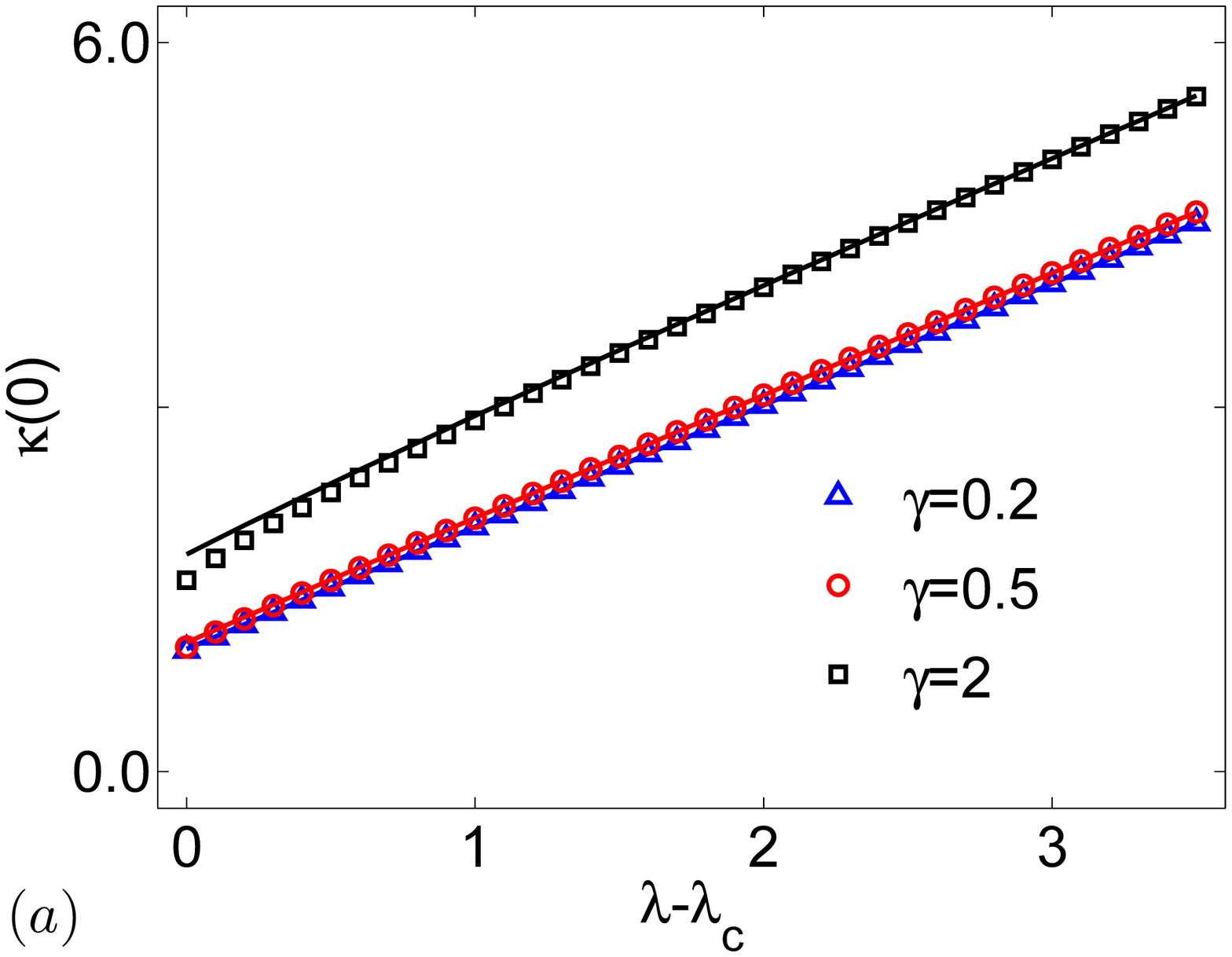} %
\includegraphics[ bb=24 178 542 584, width=6.5 cm, clip]{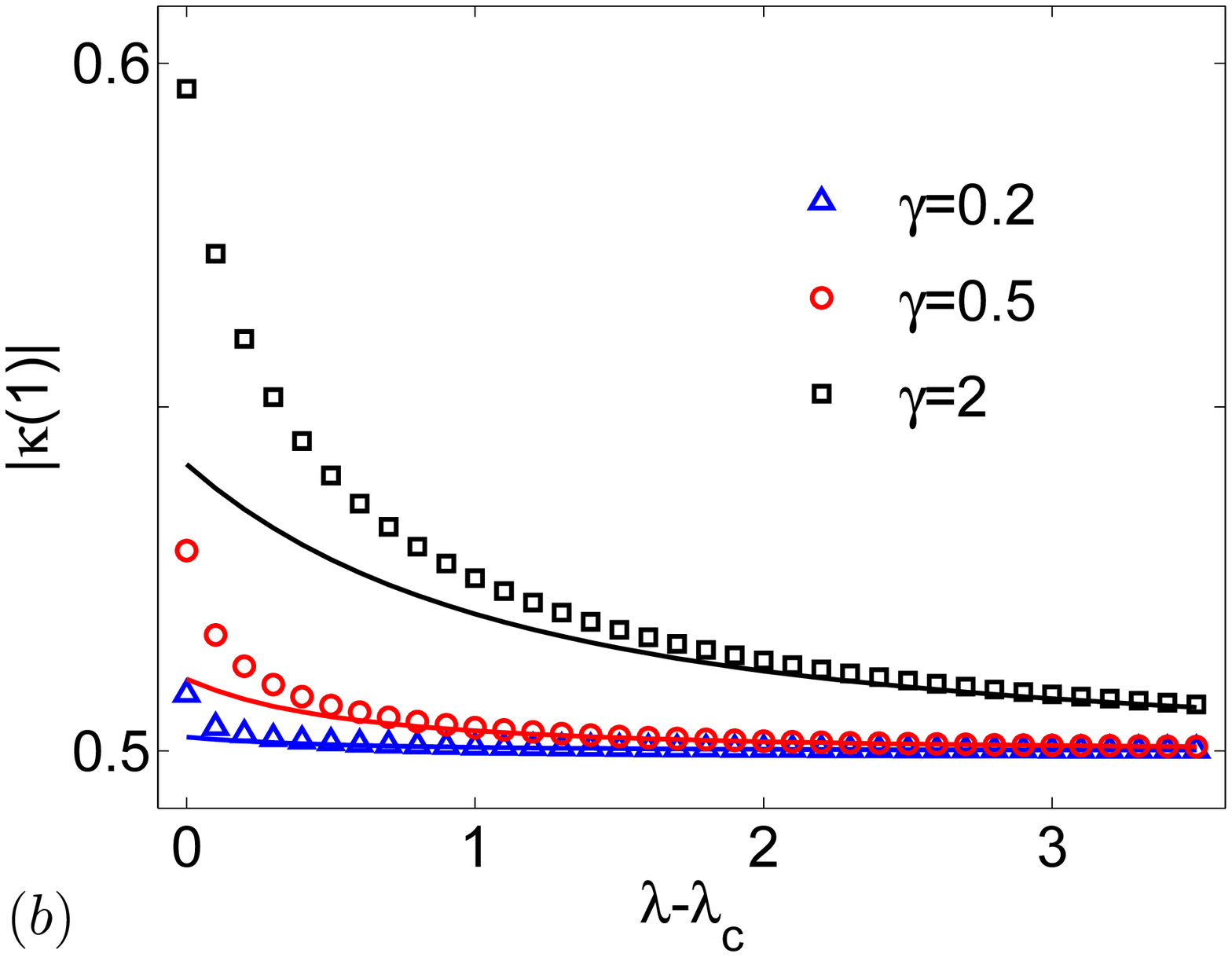} %
\includegraphics[ bb=24.5 178 542 584, width=6.5 cm, clip]{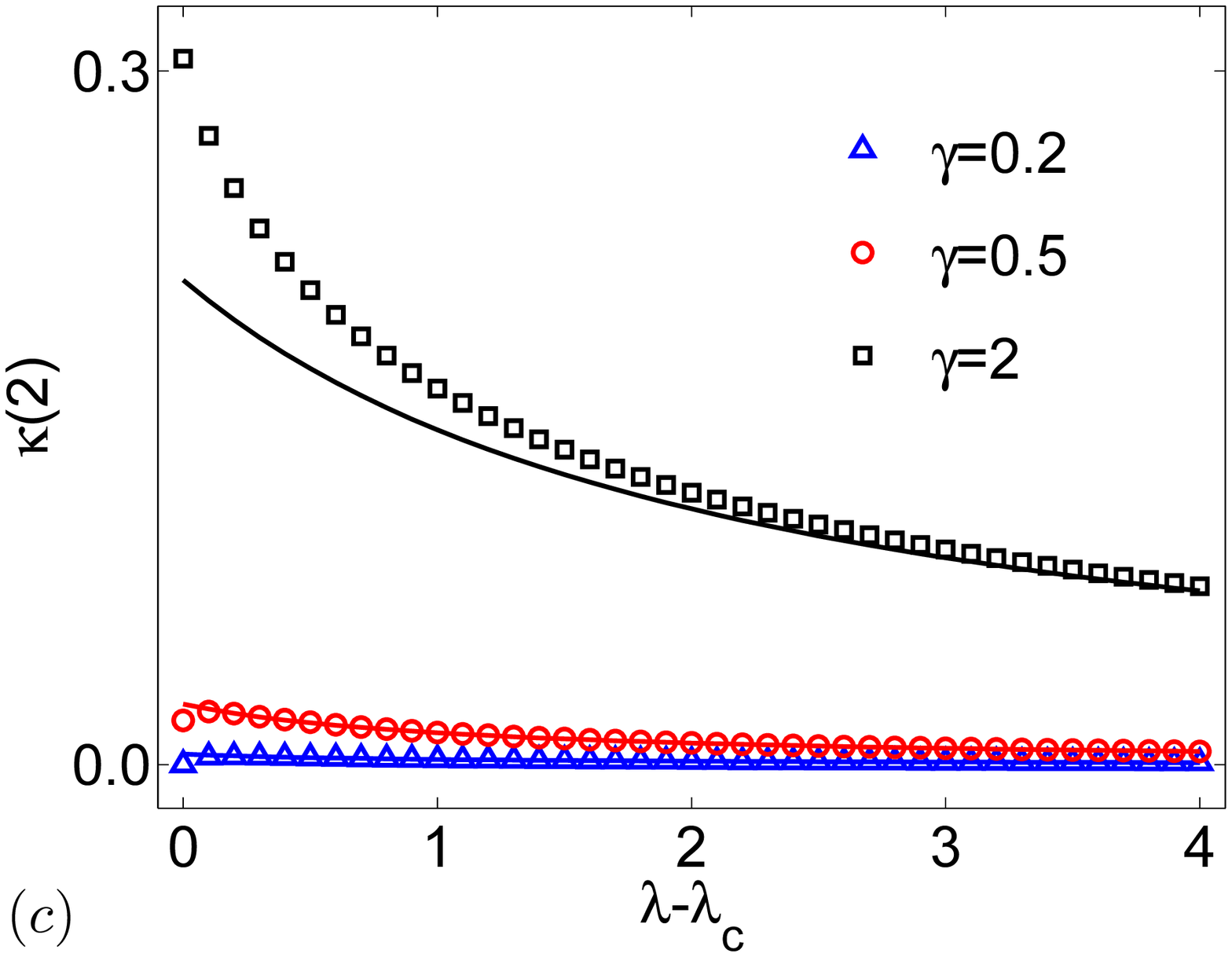} %
\includegraphics[ bb=24.5 178 542 584, width=6.5 cm, clip]{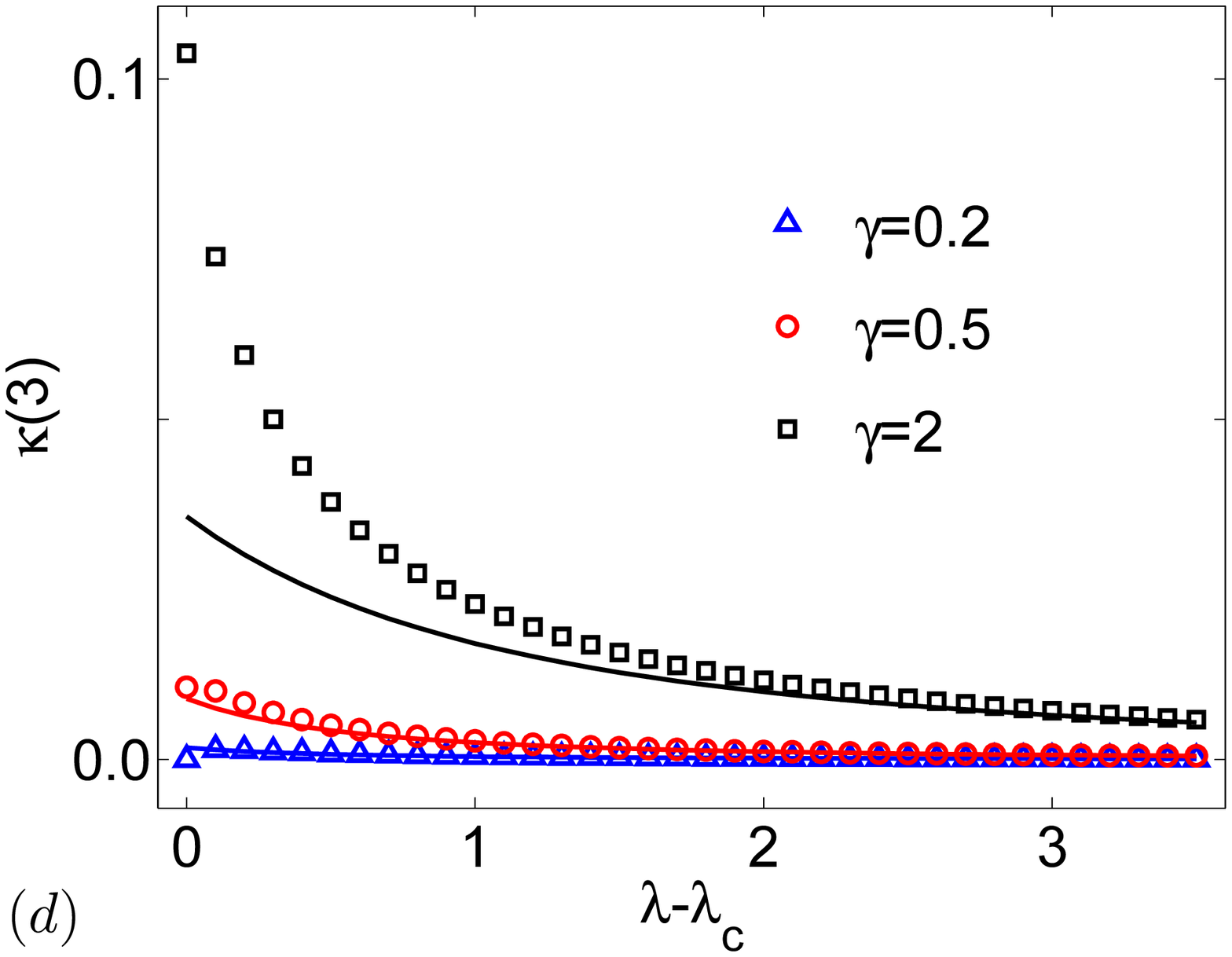}
\caption{(Color online) Plots of the coupling constant ($a$) $\protect\kappa %
\left( 0\right) $, ($b$) $\protect\kappa \left( 1\right) $, ($c$) $\protect%
\kappa \left( 2\right) $ and ($d$) $\protect\kappa \left( 3\right) $ for the
system of $N=10$. The blue solid line (empty triangle), red solid line
(empty circle) and black solid line (empty square) indicates the
approximately analytical (\protect\ref{kappa_n}) (numerical (\protect\ref%
{kappa})) result for $\protect\gamma =0.2$, $\protect\gamma =0.5$ and $%
\protect\gamma =2$, respectively. It is shown that the approximately
analytical results are in agreement with the numerical results as $\protect%
\lambda $ increases. Note that the scales of the subfigures for various
coupling constants $\protect\kappa $ are different.}
\label{fig2}
\end{figure*}


Secondly, according to the non-Hermitian quantum theory, the occurence of
the exceptional point always accomplishes the $\mathcal{RT}$\ symmetry
breaking of an eigenstate. For the concerned model, the symmetry of the
groundstate $\left\vert \text{G}_{\eta }\right\rangle $ can be indicator of
the phase transition due to the fact that the groundstate energy becomes
complex once the system is in broken region. In the following, we focus on
the discussion about the symmetry of $\left\vert \text{G}_{\eta
}\right\rangle $\ in the different regions.

Taking the combination of the Jordan-Wigner and Fourier transformations on
the rotational operator in Eq. (\ref{R}), we have

\begin{equation}
\mathcal{R}\mathcal{=}\left( -i\right) ^{N/2}\prod_{\eta =\pm ,k_{\eta }}%
\left[ 1-\sqrt{2}e^{-i\pi /4}n_{k_{\eta }}\right] ,
\end{equation}%
where $n_{k_{\eta }}=c_{k_{\eta }}^{\dag }c_{k_{\eta }}$ is the particle
number in $k_{\eta }$ space. Applying the $\mathcal{RT}$\ operator on the
fermion operators and its vacuum state $\left\vert \text{Vac}\right\rangle $%
, we have%
\begin{eqnarray}
\mathcal{RT}c_{k_{\eta }}^{\dagger }\left( \mathcal{RT}\right) ^{-1}
&=&ic_{-k_{\eta }}^{\dagger }, \\
\mathcal{RT}c_{k_{\eta }}\left( \mathcal{RT}\right) ^{-1} &=&-ic_{-k_{\eta
}},  \notag
\end{eqnarray}%
and

\begin{equation}
\mathcal{RT}\left\vert \text{Vac}\right\rangle =\left\vert \text{Vac}%
\right\rangle ,
\end{equation}%
which are available in the both regions. However, the coefficients $\cos
\left( \theta /2\right) $ and $\sin \left( \theta /2\right) $\ experience a
transition as following when the corresponding single-particle level changes
from real to imaginary: We have $\left[ \cos \left( \theta /2\right) \right]
^{\ast }$ $=\cos \left( \theta /2\right) $ and $\left[ \sin \left( \theta
/2\right) \right] ^{\ast }$ $=-\sin \left( \theta /2\right) $ for real
levels and $\left[ \cos \left( \theta /2\right) \right] ^{\ast }$ $=\sin
\left( \theta /2\right) $ for the imaginary levels, respectively. This leads
to the conclusion that the groundstate is not $\mathcal{RT}$ symmetric in
the broken region, i.e.,%
\begin{equation}
\left\{
\begin{array}{cc}
\mathcal{RT}\left\vert \text{G}_{\eta }\right\rangle =\left\vert \text{G}%
_{\eta }\right\rangle ; & \text{Unbroken} \\
\mathcal{RT}\left\vert \text{G}_{\eta }\right\rangle \neq \left\vert \text{G}%
_{\eta }\right\rangle , & \text{Broken}%
\end{array}%
\right. .
\end{equation}%
It shows that the $\mathcal{RT}$\ symmetry in the present model plays the
same role as $\mathcal{PT}$\ symmetry\ in the $\mathcal{PT}$\
pseudo-Hermitian system.

As a comparison, it is noted that the phase boundary of the Hermitian
anisotropic $XY$\ model is an ellipse. It is worth pointing out that the
phase transitions in the Hermitian and the non-Hermitian models are the
different types of quantum phase transition. The typical quantum phase
transition \cite{S. Sachdev} describes an abrupt change in the ground state
of a many-body system. For the Hermitian anisotropic $XY$ model, the
transition occurs when the order parameter, the energy gap between the
ground and first excited states, goes to zero. For the present non-Hermitian
model, zero gap also leads to the phase boundary of the ground state, which
tends to $\lambda _{c}^{2}-\gamma _{c}^{2}=1$ in thermodynamic limit. It
accords with the boundary of the unbroken $\mathcal{RT}$ symmetric region.
It reveals two differences between the phase transitions in the
non-Hermitian and the Hermitian $XY$ model: (i) The energy gap can vanish%
\textbf{\ }for the former with the finite $N$, while for the Hermitian $XY$
model zero-gap is never achieved unless the thermodynamic limit is reached.
(ii) There is no $\mathcal{RT}$ symmetry\ breaking in the quantum phase
transition of the Hermitian $XY$ model.

\section{Hermitian Counterpart}

\label{sec Hermitian counterpart}In the complex quantum theory, as an
important perspective, the physical meaning of a non-Hermitian Hamiltonian
has received a lot of attentions. When speaking of the physical significance
of a non-Hermitian Hamiltonian, one of the ways is to seek its Hermitian
counterparts \cite{A.M38}, which possess the identical real\ spectrum.
According to the complex quantum mechanics, a non-Hermitian Hamiltonian can
be transformed into a Hermitian Hamiltonian by introducing a metric, a
bounded positive-definite Hermitian operator \cite{A.M37}. However, the
obtained equivalent Hermitian Hamiltonian is usually quite complicated \cite%
{A.M37,Jin Liang}. On the one hand, it is tough to provide an explicit
mapping of a pseudo-Hermitian Hamiltonian to its equivalent Hermitian
Hamiltonian and on the other hand, the obtained Hermitian counterpart is
unphysical, which usually contains the nonlocal interactions \cite{A.M37,Jin
Liang}. In this section, we will provide a physical counterpart of the
present model and analyze its behavior when approaches to the exceptional
point.

A Hermitian Hamiltonian becomes a Hermitian counterpart of a non-Hermitian
system if and only if they share the identical real spectrum. In this sense,
the counterpart is not unique and seems to be easy constructed. However, it
is expected that the obtained counterpart is\textbf{\ }physically relevant,
or possessing the local interaction, which makes things difficult. To this
end, our strategy is to consider the operator as the form
\begin{equation}
h_{\eta }=\sum\limits_{k_{\eta }}\epsilon \left( \lambda ,k_{\eta },\gamma
\right) \left( d_{k_{\eta }}^{\dag }d_{k_{\eta }}-\frac{1}{2}\right) ,
\end{equation}%
where $d_{k_{\eta }}$ is Fermion operator satisfying%
\begin{eqnarray}
\left\{ d_{k_{\eta }}^{\dag },d_{k_{\eta }^{\prime }}\right\} &=&\delta
_{k_{\eta },k_{\eta }^{\prime }}, \\
\left\{ d_{k_{\eta }},d_{k_{\eta }^{\prime }}\right\} &=&\left\{ d_{k_{\eta
}}^{\dag },d_{k_{\eta }^{\prime }}^{\dag }\right\} =0.  \notag
\end{eqnarray}%
Here $h_{+}$ and $h_{-}$\ represent the Hamiltonians in the invariant
subspaces with even number of $\sum\nolimits_{k_{+}}d_{k_{+}}^{\dag
}d_{k_{+}}$\ and odd number of $\sum\nolimits_{k_{-}}d_{k_{-}}^{\dag
}d_{k_{-}}$, respectively. Obviously, $h_{\eta }$ is Hermitian when $%
\epsilon \left( \lambda ,k_{\eta },\gamma \right) $\ is real and has the
identical spectrum with that of $H_{\eta }$.

We employ the Fourier transformation
\begin{equation}
d_{k_{\eta }}=\frac{1}{\sqrt{N}}\sum\limits_{j}d_{j}e^{-ik_{\eta }j},
\end{equation}%
and Jordan-Wigner transformation \cite{P. Jordan}%
\begin{equation}
d_{j}=\frac{1}{2}\prod\limits_{l<j}\left( \tau _{l}^{z}\right) \tau _{j}^{+},
\end{equation}%
to rewrite the Hamiltonian $h_{\eta }$\ by the Pauli spin operators $\tau
_{l}^{\alpha }$ ($\alpha =x,$ $y,$ $z$). Here $\left\{ \tau _{l}^{\alpha
}\right\} $ and $\left\{ \sigma _{l}^{\alpha }\right\} $\ are different sets
of spin operators, which have no direct relation, or more precisely, there
is no explicit transformation to connect them.\ The essence is that we
replace the pair of conjugate operators $\left( \overline{A}_{k_{\eta
}},A_{k_{\eta }^{\prime }}\right) $ by $\left( d_{k_{\eta }}^{\dag
},d_{k_{\eta }^{\prime }}\right) $, which ensures the identical spectrum of
two Hamiltonians. However, they are distinct due to the fact that $\left(
\overline{A}_{k_{\eta }}\right) ^{\dag }\neq A_{k_{\eta }}$.

Now we transform the Hamiltonians $h_{\eta }$\ into the following spin model%
\begin{equation}
h_{\eta }=-\frac{1}{2}\sum\limits_{m>l}\kappa _{lm}\prod\limits_{l<j<m}\tau
_{j}^{z}\tau _{m}^{+}\tau _{l}^{-}+\text{H.c.}+\sum\limits_{l}\kappa
_{ll}\tau _{l}^{z},  \label{H_non_eta}
\end{equation}%
where
\begin{eqnarray}
\kappa _{lm} &=&\frac{J}{N}\sum\limits_{k_{\eta }}\sqrt{\left( \lambda -\cos
k_{\eta }\right) ^{2}-\gamma ^{2}\sin ^{2}k_{\eta }}\cos \left[ k_{\eta
}\left( l-m\right) \right]  \notag \\
&=&\kappa _{\eta }\left( l-m\right)  \label{kappa}
\end{eqnarray}%
is $\eta $-dependent coupling constant. Here $h_{+}$ and $h_{-}$\ represent
the Hamiltonians in the invariant subspaces with even and odd number of
magnons. It looks complicated due to the long-range coupling and the extra
phase $\prod\nolimits_{l<j<m}\tau _{j}^{z}$. We will show that it can be
reduced approximately to a simple model in the non-trivial parameter region.

First of all, we consider the strong filed limit case, $\lambda \gg \sqrt{%
1+\gamma ^{2}}$. Taking the Taylor expansion, we get

\begin{eqnarray}
&&\kappa _{\eta }\left( \left\vert l-m\right\vert \right) \approx \frac{J}{N}%
\sum\limits_{k_{\eta }}[\left( \lambda -\cos k_{\eta }\right) \\
&&-\frac{\left( \lambda +\cos k_{\eta }\right) \gamma ^{2}\sin ^{2}k_{\eta }%
}{2\lambda ^{2}}]e^{i\left( \left\vert l-m\right\vert \right) k_{\eta }}.
\notag
\end{eqnarray}%
Using the relation%
\begin{equation}
\sum\limits_{k_{\eta }}e^{-ik_{\eta }\left( l-m\right) }=N\delta _{l,m},
\end{equation}%
we have%
\begin{equation}
\kappa _{\eta }\left( \left\vert l-m\right\vert \right)
=J\sum\limits_{n=0}^{3}\left[ \kappa \left( n\right) \delta _{\left\vert
l-m\right\vert ,n}+O\left( \nu ^{2}\right) \right]
\end{equation}%
where $\nu =\gamma ^{2}/\lambda ^{2}$ and%
\begin{eqnarray}
\kappa \left( 0\right) &=&\left( 1-\frac{\nu }{4}\right) \lambda J,
\label{kappa_n} \\
\kappa \left( 1\right) &=&-\frac{J}{2}\left( 1+\frac{\nu }{8}\right) ,
\notag \\
\kappa \left( 2\right) &=&\frac{\nu \lambda J}{8}\text{, }\kappa \left(
3\right) =\frac{\nu J}{16}.  \notag
\end{eqnarray}%
It is shown that $\kappa _{\eta }\left( l-m\right) $ is $N$ and $\eta $%
-independent, and decays as $\left\vert l-m\right\vert $\ increases.
Secondly, numerical simulation indicates that even the field $\lambda $\ is
not so strong, $\kappa \left( l-m\right) $\ is still negligible for $%
\left\vert l-m\right\vert >1$. Fig. \ref{fig2} is the plot of $\kappa \left(
l-m\right) $ in Eq. (\ref{kappa}).

\begin{figure}[tbp]
\includegraphics[ bb=17 217 544 610, width=6.5 cm, clip]{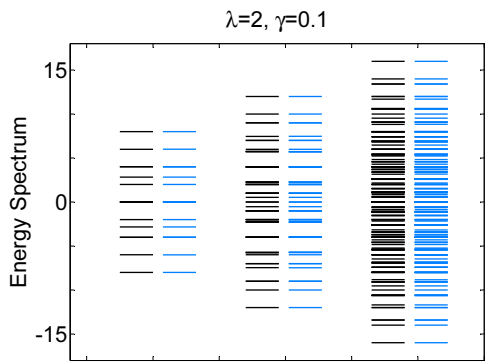}
\caption{(Color online) Plots of the energy spectrum for the systems of $%
N=4, $ $6,$ $8$\ with $\protect\lambda =2$, $\protect\gamma =0.1$. The black
and azure solid lines denotes the energy spectrum of Hamiltonian $H $ and $%
\mathcal{H}$, respectively. It can be observed that the energy spectra
obtained by two Hamiltonians are in well agreement with each other.}
\label{fig3}
\end{figure}

Then the Hamiltonian can be reduced approximately as

\begin{equation}
\mathcal{H}=-\frac{\kappa \left( 1\right) }{2}\sum\limits_{l}\left( \tau
_{l}^{+}\tau _{l+1}^{-}+\text{H.c.}\right) +\kappa \left( 0\right)
\sum\limits_{l}\tau _{l}^{z},  \label{non_Hxy}
\end{equation}%
which is the ordinary $XY$ model. To demonstrate this equivalent, we make a
comparison of $\mathcal{H}$\ and the original Hamiltonian $H$ in Eq. (\ref{H}%
).\ The energy levels are computed for small size systems. In Fig. \ref{fig3}
we see that they are in agreement well in the physical parameter region.

Secondly, it is worthy to point out that $\mathcal{H}$\ does not contain the
information of the exceptional point if we take the approximate form of $%
\kappa \left( 0\right) $ and $\kappa \left( 1\right) $ in Eq. (\ref{kappa_n}%
). As well known, distinguished from phase transition in a Hermitian system,
the non-analytic properties for the phase transition of a non-Hermitian are
caused by the Hamiltonian becoming a Jordan block operator at the
exceptional point. Nevertheless, the Jordan block cannot appear in a
Hermitian Hamiltonian. Therefore it is interesting to see what happens to
the equivalent Hamiltonian at the exceptional point. To investigate the
critical behaviors of the equivalent Hamiltonian, the exact coupling
constants $\kappa _{\eta }\left( l-m\right) $\ should be considered. We
focus on the feature of $\kappa _{\eta }\left( l-m\right) $ as $\left(
\lambda ,\gamma \right) $ turns to the critical point $\left( \lambda
_{c},\gamma _{c}\right) $ except the case of $\lambda _{c}=1$ $\left( \gamma
_{c}=0\right) $, which is not the exceptional point.\ To this end, we take
the gradient of $\kappa _{\eta }\left( l-m\right) $ in the $\lambda -\gamma $
plane, i.e.,%
\begin{eqnarray}
\nabla \kappa _{\eta } &=&\frac{\partial \kappa _{\eta }}{\partial \lambda }%
\hat{e}_{1}+\frac{\partial \kappa _{\eta }}{\partial \gamma }\hat{e}_{2} \\
&=&\frac{J}{N}\sum\limits_{k_{\eta }}\cos \left[ k_{\eta }\left( l-m\right) %
\right] \frac{\left( \lambda -\cos k_{\eta }\right) \hat{e}_{1}-\gamma \sin
^{2}k_{\eta }\hat{e}_{2}}{\sqrt{\left( \lambda -\cos k_{\eta }\right)
^{2}-\gamma ^{2}\sin ^{2}k_{\eta }}},  \notag
\end{eqnarray}%
where $\hat{e}_{1}$ and $\hat{e}_{2}\ $denote the unit vectors,$\ \hat{e}%
_{i}\cdot \hat{e}_{j}=\delta _{ij}$. When $\left( \lambda ,\gamma \right) $
turns to the critical point $\left( \lambda _{c},\gamma _{c}\right) $, the
dominant contribution to the summation is the term of $k_{\eta }=k_{c}$,
where
\begin{equation}
\cos k_{c}=\frac{1}{\lambda _{c}}.
\end{equation}%
Then we have

\begin{eqnarray}
\lim_{\left( \lambda ,\gamma \right) \rightarrow \left( \lambda _{c},\gamma
_{c}\right) }\nabla \kappa _{\eta }=\lim_{k_{\eta }\rightarrow k_{c}}\frac{J%
}{N}\cos \left[ k_{\eta }\left( l-m\right) \right] && \\
\times \frac{\left( \lambda _{c}-\cos k_{\eta }\right) \hat{e}_{1}-\gamma
_{c}\sin ^{2}k_{\eta }\hat{e}_{2}}{\sqrt{\left( \lambda _{c}-\cos k_{\eta
}\right) ^{2}-\gamma _{c}^{2}\sin ^{2}k_{\eta }}}=\infty . &&  \notag
\end{eqnarray}%
It accords with the fact that the derivative of the energy diverges when the
system tends to the exceptional point. Then for a Hermitian matrix, the
non-analytic behavior is not from the Jordan block but from the divergence
of derivatives of matrix elements. The similar situation also occurs in
another model \cite{X. Z. Zhang}, which may imply a general conclusion.

\section{Summary}

\label{sec_summary}In summary, we have proposed a non-Hermitian model
without $\mathcal{PT}$ symmetry\ explicitly, but with intrinsic $\mathcal{RT}
$ symmetry, which is a non-Hermitian version of the anisotropic $XY$ model.
Based on the exact solution, we have found the phase diagram for the finite $%
N$ system and the corresponding symmetry breaking in ground state. An
analysis of the symmetry in the ground state shows that the $\mathcal{RT}$\
symmetry is broken when the groundstate energy becomes complex, which is
similar to that for a $\mathcal{PT}$ system. It indicates that the $\mathcal{%
PT}$ symmetric model is not the unique candidate of pseudo-Hermitian system.
We also constructed a Hermitian counterpart which exhibiting the identical
real spectrum to that of the original one within the unbroken region. It is
an isotropic $XY$\ spin chain with long-range coupling, which has been shown
to have the following properties: The Hermitian counterpart can be reduced
to a simple $XY$ model with nearest neighbor coupling when the system is not
in the vicinity of the exceptional point. The derivatives of all the
coupling strengths with respect to the system parameters diverge at the
critical points. This provides an interpretation for the critical behavior
of level repulsion.\ The result for such a concrete example may have
profound theoretical and methodological implications.

\acknowledgments We acknowledge the support of the National Basic Research
Program (973 Program) of China under Grant No. 2012CB921900.

\end{document}